\documentclass[aip,rsi,amsmath,amssymb,twocolumn,reprint]{revtex4-1}
\usepackage{graphicx}
\usepackage{dcolumn}
\usepackage{bm}
\usepackage{color}
\usepackage{pifont}

\begin{document}
\title{A high dynamic range data acquisition system for \\a solid-state electron Electric Dipole Moment experiment}
\author{Young Jin Kim}
\author{Brandon Kunkler}
\author{Chen-Yu Liu}
\author{Gerard Visser}
\affiliation{CEEM, Physics Department, Indiana University, Bloomington, IN 47408, USA}

\date{\today}

\begin{abstract}
We have built a high precision (24-bit) data acquisition (DAQ) system with eight simultaneously sampling input channels for the measurement of the electric dipole moment (EDM) of the electron. The DAQ system consists of two main components, a master board and eight individual analog-to-digital converter (ADC) boards. This custom DAQ system provides galvanic isolation, with fiber optic communication, between the master board and each ADC board to reduce the possibility of ground loop pickups. In addition, each ADC board is enclosed in its own heavy-duty radio frequency shielding enclosure and powered by DC batteries, to attain the ultimate low levels of channel cross-talk. In this paper, we describe the implementation of the DAQ system and scrutinize its performance.
\end{abstract}

\maketitle

\section{Introduction}
We are carrying out an experiment to measure the electric dipole moment (EDM) of the electron to test the conservation of the time-reversal symmetry assumed in the fundamental theory that governs the electron. This EDM search employs a solid-state technique using a Gadolinium Gallium Garnet (GGG, Gd$_3$Ga$_5$O$_{12}$) paramagnetic insulator at low temperatures~\cite{lamo,liu}. The experiment aims to measure a small magnetic field ($\sim$fT) generated by the Stark-induced magnetization in the solid through the unique EDM coupling to an external electric field. In our current experimental setup, a superconducting quantum interference device (SQUID) is used as the magnetometer to monitor the magnetic signal, as the polarity of the applied electric field is modulated at a frequency of a few Hz. At 10~mK, the spin alignment in the GGG sample is enhanced as the thermal fluctuation is reduced, and we would expect an induced magnetic flux of 17~$\mu\Phi_{0}$ with an applied electric field of 10~kV/cm, if the EDM of each unpaired electron in the solid was as large as $10^{-27} $e$\cdot$cm~\cite{lamo}. Here the flux quanta $\Phi_{0}=2.07\times10^{-7}$G$\cdot$cm$^2$. This EDM-induced magnetic flux is large enough to be measured using a standard DC SQUID magnetometer operated at 4~K. With a typical SQUID transfer function of 2~V/$\Phi_0$ and a $\sim$1~\% flux coupling efficiency from the sample to the magnetometer, we expect a voltage signal output from the SQUID electronics to be about 340~nV, without further amplifications. A typical data acquisition (DAQ) system with a 16-bit resolution (i.e., 0.3~mV resolution in $\pm$10~V input range) is not sufficient to measure such a small voltage signal. In addition, the experiment requires simultaneous sampling of voltage signals from the magnetometer, high voltage monitors, and leakage current monitors, each with a very different voltage scale. To meet these stringent requirements, we developed an ultra-high precision 24-bit DAQ system with eight input channels for simultaneously sampling the analog voltage signals of interest.

Because the expected EDM-induced magnetic signal to be measured by the SQUID sensor is very small, any possible signal contamination from other voltage monitoring channels through capacitive coupling is intolerable. In particular, the high voltage monitoring channels have very large voltage that is in phase with the magnetization signal. To address this problem, that has been plaguing the experiment since the very beginning, we take extra efforts to design a custom DAQ system to have each of the analog input channels individually shielded in its own isolated heavy-duty radio frequency (RF) shielding enclosure, with galvanic isolation from the rest of the system. Fiber optic communications to the master board are used for the control of the measurement sequences and the retrieval of the digitized data. With these features, the DAQ system is expected to minimize cross-talk between channels, reduce electromagnetic interference, and eliminate the possibility of unwanted currents flowing in the ground loops that result in increased noise. Finally, to reduce the random noise and reach the desired EDM sensitivity, we need to repeat the EDM experiment over many polarity modulation cycles and carry out the average of the accumulated data sets. Therefore, ensuring that the DAQ system has no sources of non-Gaussian noise at the level of the required voltage sensitivity is essential for the success of the EDM experiment. A system capable of such performance requirements is currently not commercially available.

Our custom DAQ system can be used by other experiments that require simultaneous monitoring of several voltage sources. The system can accommodate both very low-level and large-amplitude signals with a large dynamic range that comes with a 24-bit resolution. More importantly, much care is given to ensure good rejection of spurious couplings between channels, through individual shielding of each input channel with dedicated ADC board and reduction of ground loops through optical communications. We explain the hardware and software of the DAQ system in Sec.~\ref{2} and evaluate its overall performance in Sec.~\ref{3}.

\section{Hardware design}\label{2}
\begin{figure}[t]
\centering
\includegraphics{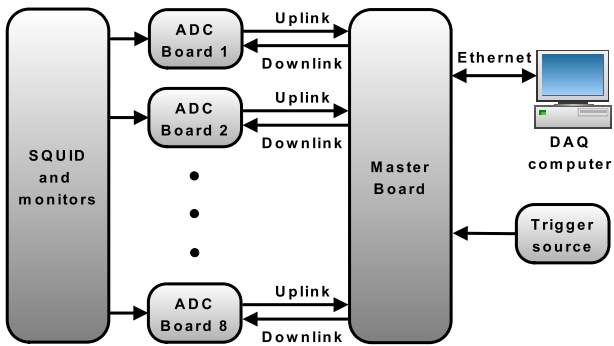}
\caption{ \label{daq} Block diagram of the DAQ system.}
\end{figure}

As shown in Fig.~\ref{daq}, the DAQ system has a master board to control eight independent modular ADC boards, each containing a 24-bit ADC chip and supporting electronic components. The front-end of the ADC boards connects to the various analog voltage sources in the experiment that need to be measured, digitized, and recorded. The ADC boards can be placed as close as possible to the experiment, with long optical fibers transmitting the digitized signals back to the master board for temporary data buffering. The master board is equipped with a FPGA chip that can be programmed for specific tasks and DAQ sequences. The communication between the master board and the ADC boards is implemented through serial fiber optic data links in both ways. The data is sent to the DAQ computer at specified intervals. Any PC running a MATLAB program can be used as the DAQ computer to interface with the master board through an optically coupled Ethernet port. The device acquires an IP address and thus can be remotely accessed through any computer connected to the Internet. This DAQ computer provides the overall control of data acquisition, data storage and analysis. The DAQ system is triggered through an external trigger source (with TTL signals) which can be controlled independently by the DAQ computer.

In addition to the DAQ function, we have added to the system a precision 20-bit digital-to-analog converter (DAC) board, to supply a low drift analog voltage source. The DAC board will be used to drive a high voltage amplifier in the next generation EDM experiment. The individual component of the DAQ system will be discussed in details in the following.

\subsection{ADC board}\label{sec:adc}
The ADC board uses a differential-input, 24-bit delta-sigma ADC chip (LTC2440) made by Linear Technology~\cite{ADC}. Since a high dynamic range with a low noise level is the essential feature of this custom system, we paid extra attention to the noise from different parts of the system. The intrinsic noise of the ADC chip is estimated to be 200~n$V_{rms}$ when sampled at 6.9~Hz (with lots of internal oversampling). The sample rate can be increased up to 3.5~kHz at the cost of a larger noise. To preserve this noise figure, we implement the analog front-end with low-noise, precision operational amplifiers (LT1007) that have a high common-mode rejection. The voltage input can be connected single-endedly or differentially. The low degree of voltage noise and fluctuation is further ensured by the use of a very low-noise voltage reference (ADR445) with an adjustment to optimize the common-mode rejection ratio (CMRR).
\begin{figure}[t]
\centering
\includegraphics{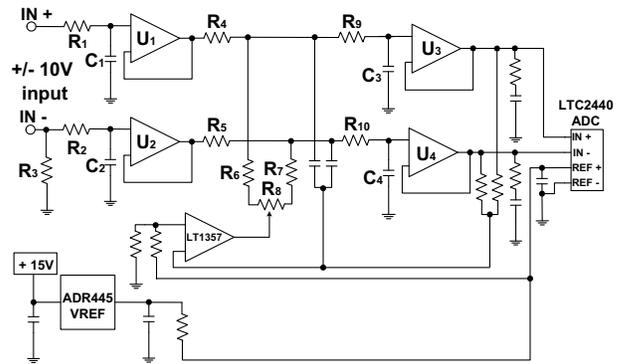}
\caption{\label{analog} Simplified schematic diagram of the low noise front-end of the ADC board.}
\end{figure}

The schematic diagram of the analog front-end is shown in Fig.~\ref{analog}. The input stage is comprised of a low pass filter (R$_1$, C$_1$, and R$_2$, C$_2$ for each differential input) to remove high frequency noises, and a unity gain buffer (U$_1$ and U$_2$) to provide a high input impedance of 7~G$\Omega$ to prevent any significant loading and perturbation of the voltage source. The resistor R$_3$ provides a bypass ground path for the operational amplifier (op-amp) bias current when the source ground is not common to the ADC Box.  The attenuation stage (R$_4$-R$_7$) linearly attenuates the $\pm$10~V input signal, the voltage swing from a typical physics experiment, to a $\pm$2.5~V signal that is compatible with the input voltage range of the ADC chip (LTC2440). The input stage is followed by the buffer stage that is comprised of an anti-aliasing filer (R$_9$, C$_3$, and R$_{10}$, C$_4$ for each differential input) and a unity gain buffer (U$_3$ and U$_4$). The anti-aliasing filter attenuates any signal with frequencies above half of the sampling frequency to prevent high frequency interferences from being shifted into the frequency band of interest during sampling. The high impedance of the buffer prevents over-loading of the attenuation stage. Finally, the feedback stage ensures the differential voltage to be centered in the ADC input range around 0~V. The CMRR adjustment is made with potentiometer R$_8$. The overall gain accuracy is approximately 0.3~\%.

\begin{figure}[t]
\centering
\includegraphics[width=3.2in,keepaspectratio]{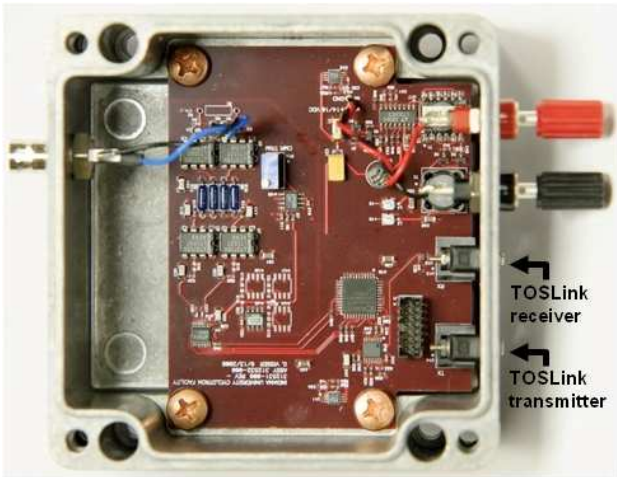}
\caption{\label{ADC} Assembled ADC board inside a heavy-duty RF shielding enclosure that is 12~cm$\times$12~cm in size. The analog voltage signal is input into the BNC connector on the left, and the 12~VDC powers is connected through the black and red banana connectors on the right. Two TOSLink optical modules (transmitter and receiver) are also on the right. }
\end{figure}

The ADC control through optical communications is implemented in a complex programmable logic device (CPLD) on the same ADC board. The ADC sample clock signal is recovered from the encoded data received over the serial optical interface (see Sec.~2.2) with a phased-locked loop (PLL). Each voltage input channel has a dedicated ADC board, that is mounted in its own metal RF shielding enclosure. Fig.~\ref{ADC} shows the photo of an assembled ADC board in the metal box, that is 12~cm$\times$12~cm in size. This ADC board is powered by a clean 12~VDC car battery supply at 110~mA to eliminate any power line frequency and the switching regulator RF interference used in most of the modern power supplies. The required $\pm$15~V and $\pm$5~V supplies for the internal circuitry are generated on-board. Note that the enclosure is connected to the zero voltage reference set by the battery. The BNC input shield is isolated from the chassis to prevent ground loops.

\subsection{Serial optical interface}
The serial optical interface between the master board and each ADC board is implemented with inexpensive TOSLink optical modules and cables, commonly used for digital audio. Each ADC board has its own pair of optical fibers that can be connected to the master board. This optical interface is used to provide galvanic isolations of the ADC boards from each other and from the master board. This feature significantly reduces the possibility of ground loop formations and spurious noise pickups. Serial communication is carried out with a custom data encoding scheme that ensures a 50~\% duty cycle, allowing its use with both the optical transmitter and receiver. The data encoding scheme also embeds the clock signal in the transmitted data so that the synchronized clock signal (sent by the master board) can be easily recovered using the low cost PLL chip on the individual ADC board.

\subsection{Master board}\label{masterboard}
\begin{figure}
\centering
\includegraphics{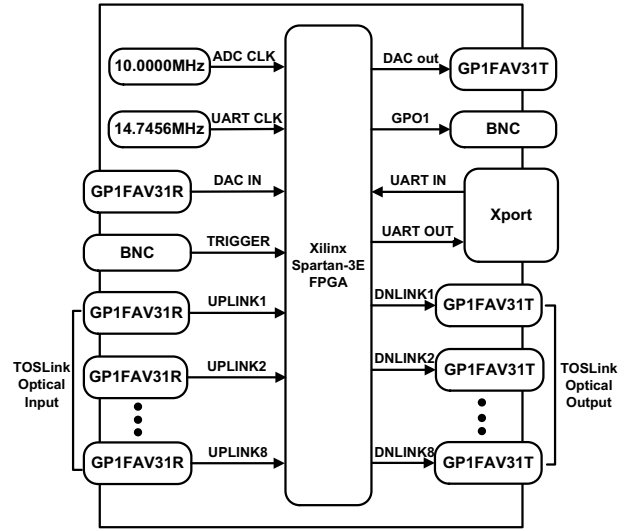}
\caption{\label{main}Block diagram of the master board.}
\end{figure}
The master board controls the DAQ sequence, renders communications with the ADC boards, implements packetization of the digitized data from ADC, and provides Ethernet connectivity with the DAQ computer. Fig.~\ref{main} shows a functional block diagram of the master board. All major functions are contained within a Spartan-3E field programmable gate array (FPGA) made by Xilinx$^{\small{\circledR}}$. The FPGA parses Ethernet packets from, and transmits Ethernet Packets to the DAQ computer. Ethernet connectivity between the computer and the FPGA is implemented with the use of the Lantronix$^{\small{\circledR}}$ XPort$^{TM}$ embedded Ethernet device server~\cite{xport} that provides a RS-232 serial port interface connected to the FPGA. Data transferred between the device server and the FPGA by way of an universal asynchronous receiver/transmitter (UART) that resides within the FPGA. All UART functions within the FPGA operate on the 14.7456~MHz clock oscillator, from which all standard baud rates can be derived. The Ethernet downlink from the DAQ computer contains data words that control the oversampling rate (OSR) of the ADC. The FPGA transmits the OSR value from the Ethernet downlink to every ADC board over the optical downlink on every rising edge of the trigger input. Every ADC samples its analog input upon receiving the OSR value. As a result, every ADC in the system reads simultaneously upon the trigger input, which defines the sample rate. Data transactions between the master board and the ADC board occur at a baud rate of 625~kbps, which is derived from the 10~MHz clock oscillator.

The CPLD on the ADC board transmits the current sample over the optical uplink each time an OSR value is received. The master board receives the ADC sample from each ADC board over the optical uplink. Once parsed, all ADC samples are multiplexed, along with a time-stamp, into a first-in-first-out (FIFO) buffer that is implemented in the FPGA. These data frames are stored in the FIFO until a request for data is made by the DAQ computer through the DAQ software (such as MATLAB), at which point it is sent to the device server by way of the RS-232 interface. Finally, the device server transmits the data frames, that contain a time stamp and sample from every ADC channel, over the Ethernet uplink to the DAQ computer.

To expand the capability of the system, the master board also contains another set of optical interface (DAC in and DAC out) and a general purpose output (GPO1). The DAC interface, which may be used to control a DAC, is identical to the ADC board interfaces; it uses the same optical connectors. The GPO1 output is capable of driving a 50~$\Omega$ load, making it useful for triggering other devices to be in sync with the DAQ system.

\subsection{DAC board}\label{sec:dac}
\begin{figure}[t]
\centering
\includegraphics{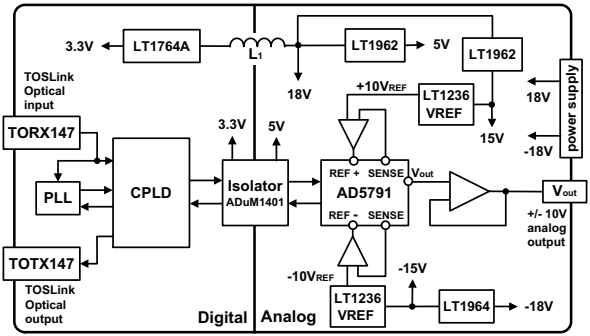}
\caption{\label{dac}Simplified schematic diagram of the precision DAC board.}
\end{figure}
We have also built a precision DAC board which can be controlled by the master board in the same way as the ADC board. A simplified schematic of the DAC board is shown in Fig.~\ref{dac}. The serial optical interface with the master board is implemented with the same optical modules (TORX147 and TOTX147) and optical fibers as the ADC board. The same PLL as the ADC board is used to recover the sample clock signal. The DAC function is accomplished with a 20-bit Analog Devices DAC chip (AD5791) and supporting components with low-noise and low temperature drift. The DAC chip offers a total harmonic distortion of $<$-97~dB, a low noise of 7.5~nV/$\sqrt{Hz}$, and a low temperature drift of $<$0.05~ppm/$^\circ$C~\cite{DAC}. This DAC chip is connected in a force sense reference configuration that uses the low-noise op-amp (AD8675) for the reference buffer and the ultra-low drift voltage references (LT1236) to apply a very stable $\pm$10~V voltage reference, in order to minimize errors caused by varying reference currents. This reference circuit is very important as the DAC output is derived from the voltage reference inputs. The force sense reference buffers are required to accurately sense and compensate a voltage drop on the reference inputs. This reference configuration provides best linearity performance of the DAC chip.

The DAC output is buffered using another AD8675 op-amp in a unity gain configuration. Because the output impedance of the DAC chip is 3.4~k$\Omega$, the output buffer is required for driving low resistive loads. The analog power (LT1962 and LT1964) and digital power (LT1764A) are isolated using the digital isolator (ADuM1401) which, along with inductor L$_1$, prevent digital noise from spreading to the analog supply. This DAC board is operated by a $\pm$18~V DC power supply. The analog output waveform from this custom DAC board is created by the DAQ computer and the output range is in $\pm$10~V.

\subsection{Data acquisition software}
The DAQ software, written in MATLAB, provides data acquisition control, data storage, and data analysis. Ethernet connectivity with the master board is achieved with the free TCP/UDP/IP toolbox. The function is implemented as a MEX-file which allows one to interface C subroutines (dynamic link libraries) to MATLAB. This DAQ software collects data from each voltage monitoring channel at a fixed trigger rate and stores the data to a disk in the DAQ computer. Functions for data analysis such as numerical average or data filtering are user implemented.

\section{Performance characterization}\label{3}
Evaluation of the performance of this custom DAQ system is necessary prior to use in the EDM experiment. The significant characteristics that need to be evaluated include the intrinsic root-mean-squared (rms) noise (without load), the cross-talk between channels, the settling time, the CMRR, the power supply rejection ratio (PSRR), and the linearity of the DAQ system.

\subsection{Intrinsic rms noise}
The intrinsic rms noise of the DAQ system stems from noise of the ADC chips and supporting circuitry.
On the level of the ADC chip, we expect the intrinsic rms noise to vary with the OSR value which defines
the effective bandwidth of the on-chip digital filter, and the voltage level in the following ways: the rms noise increases by approximately $\sqrt{2}$ when OSR decreases by a factor of 2 from OSR=32768 to OSR=256. The exception is that the rms noise at OSR=128 and OSR=64 has additional contributions from the internal modulator quantization noise (see Ref.~3). The conversion between the OSR and the sampling rate can be found in Table~\ref{osr}.
\begin{table}[b!]
\caption{\label{osr} Maximum bandwidths and ENOBs} 
    \begin{tabular}{  c  c  c  c }
    \hline\hline  
    OSR & Maximum & Measured & ADC chip \\
    \mbox{} & sampling rate & ENOB & Spec. ENOB \\
    \hline
    64 & 2.9~kHz & 18.0 &17.0   \\
    128 & 1.9~kHz & 20.4 & 20.0  \\
    256 & 976~Hz & 21.1 & 21.3  \\
    512& 488~Hz & 21.6 &  21.8 \\
    1024& 244~Hz & 22.1 & 22.4  \\
    2048& 122~Hz & 22.6 & 22.9  \\
    4096 & 61~Hz & 23.0 & 23.4  \\
    8192& 30~Hz & 23.4 & 24.0  \\
    16384& 15~Hz & 23.8 & 24.4  \\
    32768& 7~Hz & 24.1 & 24.6  \\
    \hline\hline
    \end{tabular}
\end{table}
\begin{figure}[t!]
\centering
\includegraphics{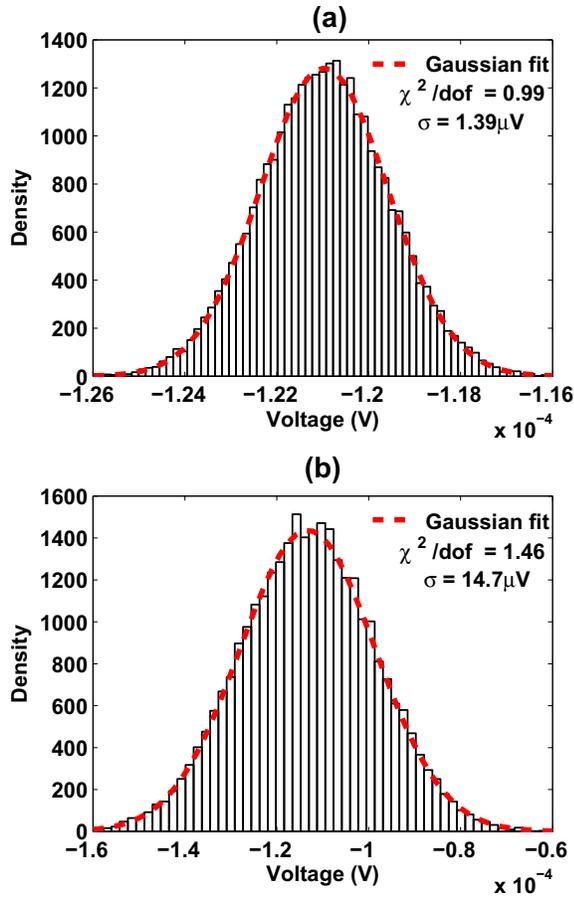}
\caption{\label{histogram} Histograms of intrinsic rms noise of the DAQ system at OSR=16384 (a) and OSR=128 (b). The analog input is terminated in these measurements. The red lines is the Gaussian fit providing the standard deviation of each histograms as the rms noise.}
\end{figure}

To assess the rms noise at different OSR values, we collected and analyzed a large number of data from the DAQ system with the analog input terminated on the ADC board under test. Fig.~\ref{histogram} shows the histogram of a typical set of data with a total of 30,000 samples, collected at OSR=16384 and OSR=128 with a sampling rate of 15~Hz and 1.5~kHz respectively. The voltage distributions can be fitted by a Gaussian function, with the standard deviation, $\sigma$, corresponding to the intrinsic rms noise of 1.39~$\mu$V and 14.7~$\mu$V at OSR=16384 and 128 respectively. These noise figures agree quite well with the noise values from the ADC chip specification~\cite{ADC}. The effective number of bits (ENOB) at OSR=16384 and 128 are measured to be 23.8 and 20.4 respectively, just slightly less than the specified values of 24.4 and 20 listed on the data-sheet of LTC2440 ADC chip. The system as a whole does not introduce significantly more noise on top of the intrinsic noise of the ADC chip.
\begin{figure}[t!]
\centering
\includegraphics{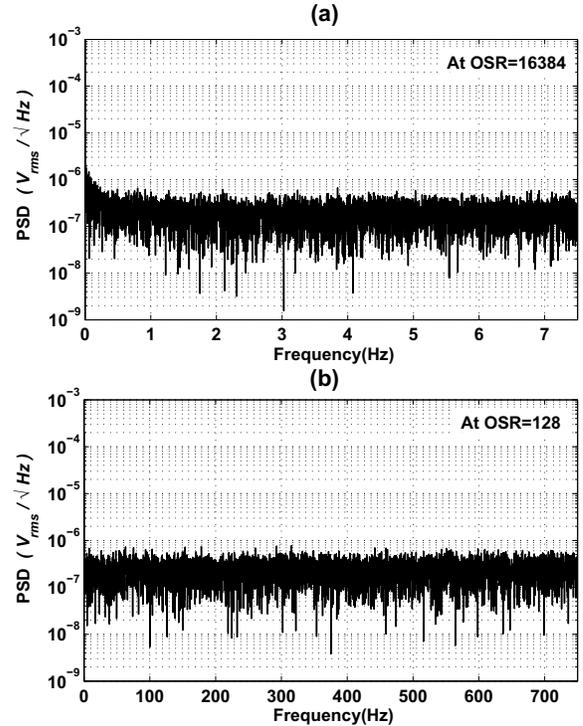}
\caption{\label{FFT}Power Spectral Density (PSD) spectra of the intrinsic rms noise of the DAQ system at OSR=16384 (a) and 128 (b). The vertical axis is in log scale.}
\end{figure}

In addition, the power spectral density (PSD) spectra of the noise measurement (shown in Fig.~\ref{FFT}) do not show any observable peaks, in particular, at the AC power supply frequency of 60~Hz and the high order harmonics.
This demonstrates that the system as a whole has no significant ground loop pickups in the DAQ system above the intrinsic noise level. More importantly, there exist no additional sources of spurious noise from trigger or digital switching that give rise to non-Gaussian noise, that can not be suppressed by taking longer average. The base level of the noise power spectrum of the whole DAQ system is measured to be 0.21~$\mu$$V_{rms}/$$\sqrt{Hz}$ at 1~Hz. Table.~\ref{osr} shows a comprehensive list of the maximum bandwidths (equal to half of maximum sampling rate) and the measured ENOBs at all available OSR values. Note that the maximum sampling rate at OSR=64 is limited by the maximum XPort$^{TM}$ baud rate of 921600~bps~\cite{xport}.
\begin{figure}[t!]
\centering
\includegraphics{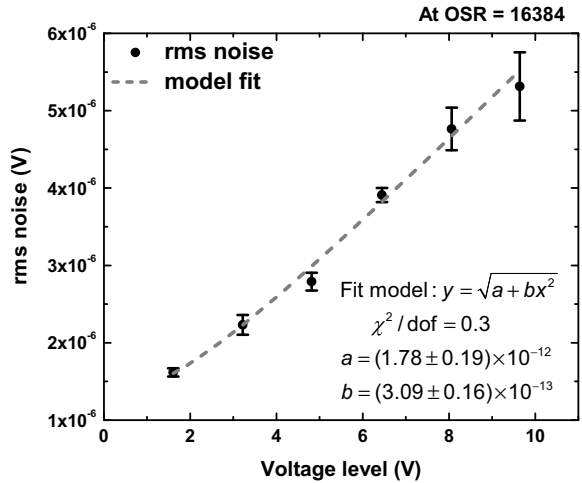}
\caption{\label{input}Intrinsic rms noise of the DAQ system as a function of the input voltage. The red curve shows the model fit.}
\end{figure}

The intrinsic rms noise of the DAQ system also depends on the voltage level of the analog input. The procedure this noise test is the same as the preceding evaluation, except that a test voltage source is connected to the analog input of the ADC board. For a low noise performance, the test voltage source is made of 1.5~V batteries (Energizer$_{\tiny{\circledR}}$ Industrial$_{\tiny{TM}}$ AA) connected in series. In practice, the battery pack has its own intrinsic noise which would be added to that of the DAQ system so that the noise test may not correctly reflect the intrinsic noise of the DAQ system. We used the normalized covariance computation~\cite{ncov} to further test whether the noise of the battery is small enough to be negligible. In the test, we utilized two ADC boards to simultaneously sample the voltage output from the same battery pack, with a sampling rate of 15~Hz and an OSR of 16384. We then estimate the extent to which the fluctuations of the data sets collected from the two ADC boards are correlated. A strong correlation would indicate that the noise contribution from the battery pack is large compared to that from the DAQ system. The analysis shows that the normalized covariance is less than 0.1, signifying weak correlation and verifies that the battery noise is negligible in these noise measurements. The measured rms noise of the DAQ system as a function of the input voltage level is plotted in Fig.~\ref{input}. The rms noise increases as the input voltage rises. The error bars are statistical and correspond to one standard deviation. We fit the functional dependence of the rms noise using $\sqrt{a+bV^2}$, where $a$ characterizes the intrinsic rms noise without the analog input, and $b$ characterizes the noise dependency of the input voltage level. The fit model was chosen to include the independent contributions from the PSD of the noise without loads and the PSD of the noise that varies with the voltage input. The result shows that the rms noise without loads (at V=0) is (1.33 $\pm$ 0.44)~$\mu$V, matching the result in the preceding noise test within the error bar.
\begin{figure}
\centering
\includegraphics{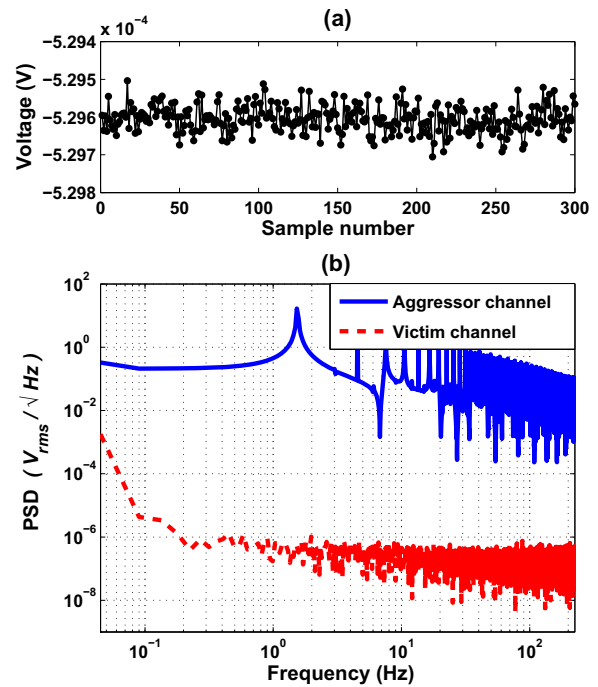}
\caption{\label{cross} Channel cross-talk measurement. (a) Digitized output of the victim channel, averaged over 41548 cycles of the square waveform. (b) PSD spectra of the aggressor channel (blue curve) and the victim channel (red curve) in log-log scale.}
\end{figure}

\subsection{Channel cross-talk}
Cross-talk between individual channels due to any feedthrough coupling, such as mutual capacitance coupling, is one of the primary systematic effects in a DAQ system. To measure the level of channel cross-talk, a 1.5~Hz square wave with an amplitude of 19~V peak-to-peak (V$_{pp}$), at a 95~\% of full scale input range, was applied to one ADC board and served as the aggressor channel. At the same time, the adjacent ADC board with the analog input terminated, as a victim channel, was sampled with an OSR of 512 and a sampling rate of 450~Hz. The digitized data from the victim channel is then averaged under the same cycle of the square waveform of the aggressor channel to reduce the random noise, thereby revealing any small contribution of the cross-talk. Fig.~\ref{cross}(a) displays the digitized signal averaged over 41548 cycles from the victim channel. The lacking of any square waveform indicates negligible cross-talk effects. The PSD spectra (Fig.~\ref{cross}(b)) of both the aggressor and the victim channels also show no measurable correlations between the two channels. In the aggressor channel, peaks at 1.5~Hz and the harmonics are evident, whereas in the victim channel no corresponding peaks are found at these frequencies. In conclusion, our custom DAQ system has a cross-talk smaller than $\sim$191~dB, much lower than any commercially available systems. The reduction of systematic effect form the channel cross-talk (in particular between the HV-monitoring and the SQUID-monitoring channels) is an indispensable requirement to accomplish the solid-state EDM experiment.
\begin{figure}
\centering
\includegraphics{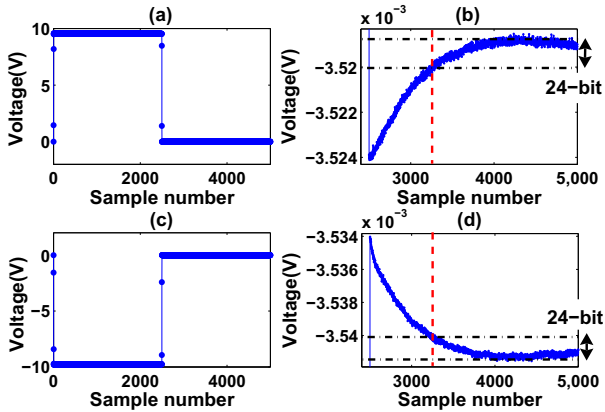}
\caption{\label{settle} Settling time measurement. (a) a step function input with from 9.5~V to 0~V, and (c) a step function from -9.5~V to 0~V. (b) and (d) are zoomed-in voltage plots around the region of voltage transition on (a) and (c). It takes 3 samples to settle the digitized output to the 22-bit resolution (see (a), (c)), and 760 samples to settle to 24-bit resolution (see (b), (d)).}
\end{figure}

\subsection{Settling time}
The settling time is defined as an elapsed time during which the output of the DAQ system settles to a desired accuracy. For an accurate EDM measurement, the settling time should be much shorter than the time period to modulate the polarity of the high voltage applied to the GGG samples.  To measure the settling time of the DAQ system, we supplied a step function as the analog voltage input to the ADC board under test. The step function should settle in a much faster time than the DAQ system. Therefore, we employed the PhotoMos relay (AQV22\ding{109}) as the test pulser to generate an instantaneous step function with high speed switching time around 0.03~ms~\cite{relay}. Two types of step input generated from the pulser are applied to the ADC board (see (a) and (c) in Fig.~\ref{settle}): one step (a) decreases from 9.5~V to zero and the other step (c) increases from -9.5~V to zero with a frequency of 3~mHz. Using the ADC board under test, we sampled the step function cycles for 200 cycles with an OSR of 16384 and a sampling rate of 15~Hz. The averaged results are shown in Fig.~\ref{settle}. Upon the voltage switch, the digitized output settles to 22-bit resolution within three samples, corresponding to a settling time of 200~ms (Fig.~\ref{settle}(a) and (c)). To settle to the 24-bit resolution, it takes 760 samples and thus a much longer time around 51~s (Fig.\ref{settle}(b) and (d)).
\begin{figure}
\centering
\includegraphics{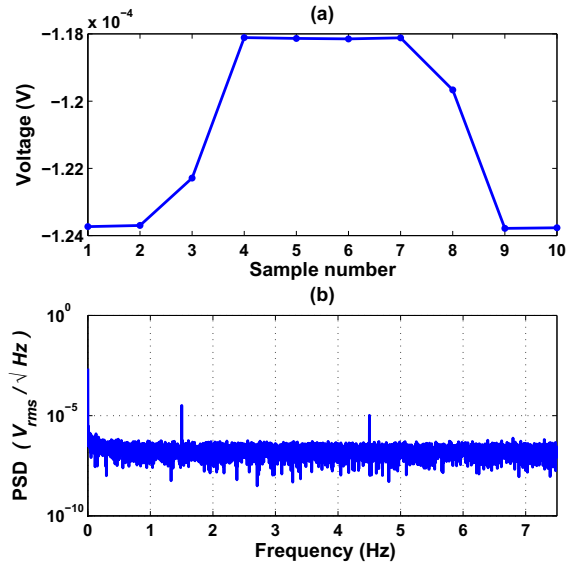}
\caption{\label{common} Common-mode rejection test. (a) Digitized output averaged over 7199 cycles. (b) The PSD spectrum of the output of 10,000 samples. The high and low analog inputs are connected to a common voltage source.}
\end{figure}

\subsection{Common-mode rejection}
Before the analog signal is sent to the ADC to be digitized, some undesirable common-mode noise (undesirably picked up from any ambient sources) is always present on both the high and low input wires of the ADC board, equal in both the phase and amplitude. This common-mode noise, quite often, is generated by capacitive couplings between the wires and ground. For the best performance, the DAQ system must suppress the common-mode noise sufficiently so as not to add additional noise, and in the mean time not to distort any voltage input of interest. To measure the CMRR, we connect both the high and low input of the ADC board under test to a common voltage source of a 1.5~Hz square waveform with a 4~V$_{pp}$ amplitude. Data was collected at OSR=16384 with a sampling rate of 15~Hz. Fig.~\ref{common}(a) shows the digitized data averaged over 7199 cycles. Notice that even with the input as a square wave, the averaged output waveform is distorted because of the discrepancies on the phase and amplitude between the high and low analog inputs, as a result of the common-mode rejection adjustment. The PSD spectrum is shown in Fig.~\ref{common}(b), where the apparent peak at 1.5~Hz (and 4.5~Hz) is measurable, which indicates some degree of CMRR. By comparing the amplitude of the output, 5.68~$\mu$V$_{pp}$, to the applied waveform strength of 4~V$_{pp}$, the CMRR is estimated to be 1~ppm, which is good enough for the EDM experiment.

\subsection{Power supply rejection}
\begin{figure}
\centering
\includegraphics{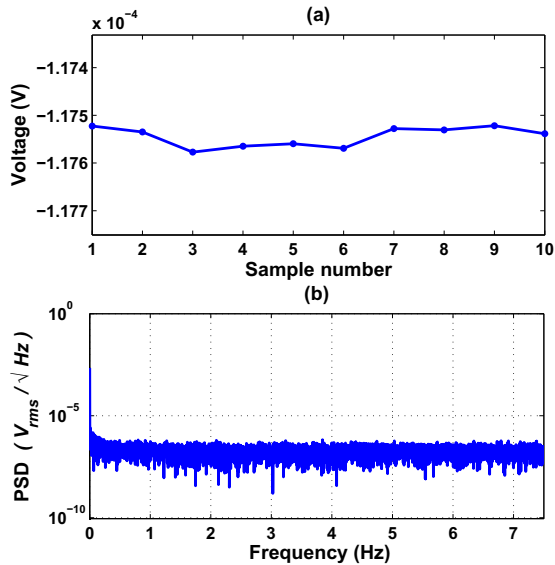}
\caption{\label{power} Power supply noise measurement. (a) Digitized data averaged over 4798 cycles. (b) The PSD spectrum of the output (with a terminated input) of 10,000 samples}
\end{figure}

Even with the most careful choice, it is inevitable that the DC power supply used to operate the ADC board has some degree of noise, such as the voltage ripples, that can affect the performance of the DAQ system. This unwanted noise from the power supply can couple parasitically through the circuitry to the analog voltage input, adding undesirable noise to the digitized outputs. We quantify the ability of the DAQ system to reject power supply noise by a) mixing a 1~V$_{pp}$, 1.5~Hz square wave together with a 12~VDC to create a ``rippled`` supply voltage, b) terminating the input of the ADC board, and c) collecting digitized data with an OSR of 16384 at a sampling rate of 15~Hz. The resulting data averaged over 4798 cycles are plotted in Fig.~\ref{power}(a) with the PSD spectrum plotted in Fig.~\ref{power}(b). The time trace of the averaged data does not show any square wave corresponding to the power supply ripple. No observable peak at the frequency of the ripple is found in the PSD spectrum, either. In summary, the PSRR of this DAQ system is quite high, and the noise from power supply is negligible even with a bad power supply with large ripples.

\subsection{Linearity}
The linearity of the DAQ system characterizes how accurately a digitized result reflects the analog input. Measurement of the linearity of a high-resolution DAC system is especially difficult due to the lacking of a good calibration source. We have tried several commercial voltage sources with sinusoidal waveform output, but found that the total harmonic distortion and phase errors are too large to be useful to carry out this test. Instead, we used the newly developed low-distortion, 20-bit precision DAC board (described in Sec.~\ref{sec:dac}) as the voltage source. An input of 0.01~Hz triangular waveform with a 19~V$_{pp}$ amplitude generated by the DAC board is fed into the ADC board under test and sampled at OSR=4096 with a sampling rate of 50~Hz. We performed a least-square line fit ($y=a+bx$) on one cycle of the digitized output, separating the ramping down half-cycle from the ramping up half-cycle. The non-linearity is quantified by the residual deviation from the ideal triangular waveform.
\begin{figure}[t]
\centering
\includegraphics{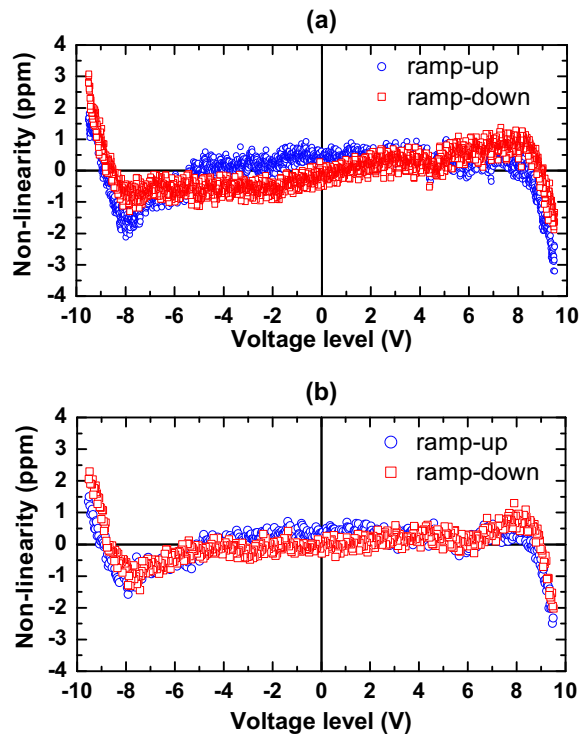}
\caption{\label{linearity}The non-linearity of the DAQ system as a function of the input voltage. (a) with a 0.01~Hz triangular waveform. (b) with a 0.04~Hz triangular waveform. The maximum non-linearity of the DAQ system is $\pm$3~ppm over the full input range.}
\end{figure}
Fig.~\ref{linearity}(a) shows the non-linearity (in ppm) as a function of the input voltage. The line fit has $\chi^2/$dof = 0.94 and 0.75 on ramp-down and ramp-up halves respectively. The maximum non-linearity of the DAQ system is estimated to be $\pm$3~ppm over the full input range ($\pm$1~ppm between voltage range of $\pm$7~V). This is good enough for the EDM experiment. The non-linearity of this DAQ system arises mainly from the LTC2440 ADC chip, with some small contributions from op-amps and resistors in the input stage and buffer stage of the ADC board (Sec.~\ref{sec:adc}). In practice, this measured non-linearity should be the combined effect from the DAC board and the DAQ system, however, without an independent calibrated voltage source we cannot isolate the non-linearity of the DAQ system. Nevertheless, the measured non-linearity is already close to the specification of the ADC chip, and thus implies that errors from the DAC board are probably insignificant. The small 1~ppm discrepancy between the ramp-up and ramp-down data sets (Fig.~\ref{linearity}(a)) is probably a result of the temperature change lagging some time behind the voltage change. Increasing the frequency of the triangular waveform reduces this discrepancy. Fig.~\ref{linearity}(b) shows the non-linearity of the DAQ system measured with the 0.04~Hz triangular waveform.

\section{Conclusion}
We have developed a high resolution 24-bit DAQ system with special attentions to the noise performance. This DAQ system is currently been used for the solid-state electron EDM experiment. In this paper, we show the detailed characterizations on the relevant parameters of the DAQ system. The measured ENOB can be as high as 24.1 when sampled at 7~Hz. The EDM measurement requires a sampling rate at $\sim$700~Hz, and the DAQ system has a ENOB of 21.1. The most important performance requirement is the ultra-low channel cross-talk which reduces the leading systematic effect observed in our EDM experiment. Using this custom DAQ system, we have not only demonstrated the feasibility of the solid state method for the electron EDM search, but also obtained the first background-free limit of the electron EDM on the order of 10$^{-25}$e$\cdot$cm ~\cite{yjk1,yjk2}.

\section{Acknowledgments}
This work was supported by NSF grants 0457219, 0758018. We also acknowledge the support from IU center for Spacetime Symmetries.

\end{document}